\documentclass[letterpaper,12pt]{article}
\renewcommand{\baselinestretch}{1.5}

\usepackage{times}
\usepackage[margin=1in]{geometry}

\usepackage{graphicx}
\DeclareGraphicsExtensions{.pdf,.jpeg,.png}
\usepackage{cite}
\usepackage{amsmath}
\usepackage{amssymb}
\usepackage{titlesec}
\titlelabel{\thetitle.\quad}
\titleformat*{\section}{\normalfont\bfseries}
\titleformat*{\subsection}{\normalfont\bfseries}
\titleformat*{\subsubsection}{\normalfont\bfseries}
\titleformat*{\paragraph}{\normalfont\bfseries}
\titleformat*{\subparagraph}{\normalfont\bfseries}

\begin{document}
\date{}

\title{Sliding Mode Control for Mixed Conventional/Braking Actuation Mobile Robots }

\author{Walelign~Nikshi\footnote{Dr. Nikshi is a Control Systems Engineer at Icon build, Austin, TX, 78745 USA e-mail: wallee394@gmail.com.},
        ~Randy~C.~Hoover\footnote{Dr. Hoover is with the Department of Electrical and Computer Engineering, South Dakota School of Mines and Technology, Rapid City, SD, 57701 USA e-mail: Randy.Hoover@sdsmt.edu.},
        ~Mark~D.~Bedillion \footnote{Dr. Bedillion is with the Department of Mechanical Engineering, Carnegie Mellon University, Pittsburgh, PA, 15213 USA e-mail: mbedillion@cmu.edu.},
        ~Saeed~Shahmiri
        \footnote {Mr. Shahmiri is Lab Coordinator and M.Sc. graduated with the Department of Electrical Engineering, South Dakota School of Mines and Technology, Rapid City, SD, 57701, USA e-mail: Saeed.Shahmiri@sdsmt.edu}
        \\and~Jeremy~Simmons \footnote{Mr. Simmons is a Ph.D student at the Department of Mechanical Engineering, University of Minnesota, Minneapolis, MN, 55455 USA e-mail: simmo536@umn.edu.}
}



\maketitle

\thispagestyle{empty}

\noindent
{\bf\normalsize Abstract}\newline
{The Mixed convention/braking Actuation Mobile Robot (MAMR) was designed to tackle some of the drawbacks of conventional mobile robots such as losing controllability due to primary actuator failures, mechanical complexity, weight, and cost. 
It replaces conventional steering wheels with braking actuators and conventional drive wheels with a single omni-directional wheel.
This makes it fall under the category of under-actuated mobile robots. 
The brakes have only two states, ON and OFF, resulting in discontinuous dynamics. This inspires the use of a discontinuous control law to control the system.
This work presents a Sliding Mode Controller (SMC) design to park the MAMR system from a given initial configuration to a desired final configuration. 
Experimental results are presented to validate the parking control of the MAMR. } \vspace{2ex}
   
\noindent
{\bf\normalsize Key Words}\newline
{Sliding mode control, conventional actuators, braking actuators, mixed conventional/braking actuators, parking control.}
\\

\section{Introduction}

{I}{n} recent years, research in the area of mobile robots has received much attention due their broad areas of application. 
Such areas include factories (e.g., automated guided vehicles used for moving parts from one point to the other), military operations (e.g., unmanned ground reconnaissance vehicles used for surveillance and monitoring), healthcare (e.g., pharmaceutical delivery), and household (e.g., floor cleaning and lawn mowing) \cite{schneier2015literature}. 
Because of the need for mobility to accomplish tasks, the dynamic nature of the environment in which the robot is operating, and the need for online modification of the robot's behavior, a special focus was given to locomotion. 
Locomotion is the primary function of mobile robots that enables them to accomplish tasks that require mobility~\cite{siegwart2011introduction}.
Mobile robots can locomote either by mimicking biological phenomena such as walking~\cite{machado2006overview, deshmukh2006robot}, skating~\cite{iverach2012ice}, swimming~\cite{dudek2005visually, dudek2007aqua}, flying~\cite{bouabdallah2007design,cutler2012design}, and others~\cite{stepan2009acroboter, wang2008biological, menon2004gecko}; or using actively powered wheels such as Wheeled Mobile Robots (WMR)~\cite{morin2008motion,de2001control}.
Locomotion schemes differ in terms of mechanical construction and control complexity.
While mobile robots can use any of these locomotion methods, locomotion systems that use wheels and electric motors are the most common technique for fairly flat and planar applications.
\begin{figure}[!t]
\centering
\includegraphics[width=\columnwidth]{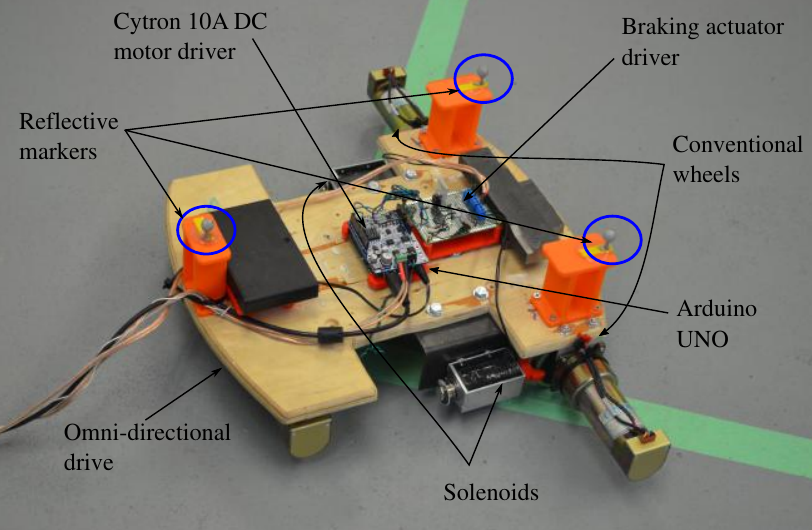}
\caption{The prototype of the MAMR used for experimental validation.
This robot is the same robot as the one used in~\cite{Walelign2018_SMC_FLC} except the Adafruit motor driver is replaced by a Cytron $10A$ DC motor driver.}
\label{MAMR_II}
\end{figure}

Mobile robots typically use conventional actuators, especially electric motors, to drive each degree of freedom. 
Such mobile robots suffer from loss of controllability if one of the conventional actuators fails during operation.
In addition, they suffer from high weight, high cost, and mechanical complexity, especially if the degrees of freedom are high.  
For example, the Sample Return Rover (SRR) from the Jet Propulsion Laboratory uses four wheels that are all powered and steered.
The SRR has been used effectively in rough environments for sample cache retrieval, lander rendezvous, and all-terrain exploration. 
However, the SRR suffers from both mechanical and control complexity as a result of each wheel being powered and steered \cite{iagnemma2000mobile,iagnemma2003control}.

\begin{figure}[!t]
\centering
\includegraphics[width=\columnwidth]{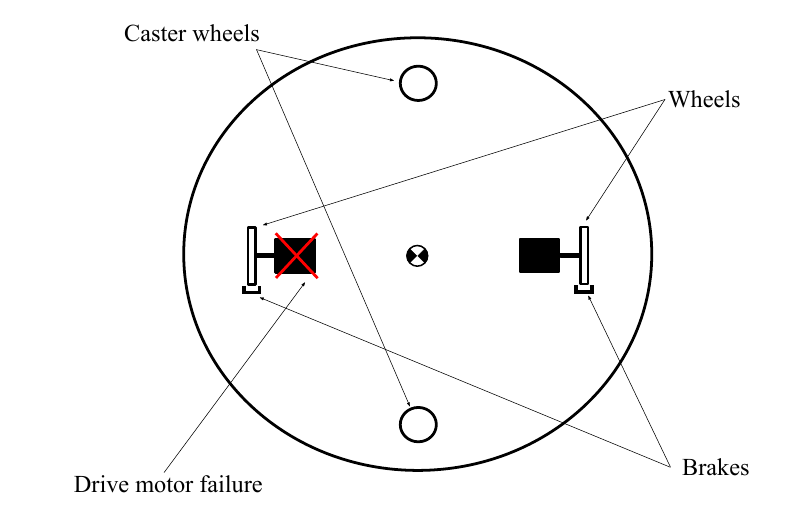}
\caption{Differential drive robot with brakes incorporated along the wheel plane \cite{Walelign2018_SMC_FLC}. }
\label{regain_controllability}
\end{figure}

To solve such problems in mobile robotics, the authors in~\cite{Walelign2017SMC,simmons2016mechatronic,nikshi2016parking,Walelign2018_SMC_FLC} proposed a new actuation approach that uses a mixture of conventional and braking actuation systems.
This approach results in a new mobile robotic platform called the Mixed conventional/braking Actuation Mobile Robot (MAMR). 
In this robot, the 
steering wheels are replaced by brakes and the conventional drive wheels by a single omni-directional drive wheel (see Fig. \ref{MAMR_II}).
The primary benefits of the MAMR platform over conventional mobile robotic platforms include \cite{Walelign2018_SMC_FLC}:
\begin{enumerate}
\item Its ability to regain the controllability of the system under actuator failure. For example, for the differential drive robot shown in Fig.~\ref{regain_controllability}, if one of the conventional drive fails during operations, the controllability of the system can be regained by activating an appropriate brake and using it as a rotation axis.
\item Its simplicity of actuation.  The use of ON/OFF brakes instead of conventional actuators can result in a simpler system in terms of actuator complexity and drive electronics.
\item  Its potential to reduce the overall weight and cost of the system.  For instance, an ON/OFF brake can be actuated by a solenoid, which can be made to be far lighter and less costly than a DC drive motor.
\item  Its potential to help in the miniaturization of mobile robots.  A brake merely requires a method of controlling the frictional contact between a robot and  its environment.  Various MEMS devices (e.g. electrostatic and thermal actuators) can be used to make and break contact with the environment or change the effective coefficient of friction.
\end{enumerate}
An additional advantage of this platform is that it can be configured for a variety of environments based on the design and type of the braking actuator.
For example, Simmons \textit{et al.}~\cite{simmons2016mechatronic} presented a prototype of the MAMR that uses a ball-type caster as a braking actuator. 
Improving on the design presented in~\cite{simmons2016mechatronic}, Nikshi \textit{et al.}~\cite{Walelign2018_SMC_FLC} presented a prototype of the MAMR that uses brakes with conventional wheels to improve the coefficient of friction and reduce the mechanical complexity due to using an omni-directional ball-type caster.

Control problems in mobile robots can be posed for various tasks such as set point stabilization, trajectory tracking, and path following \cite{solea2009sliding}. 
Because the environment in which a robot operates can be complex and inherently uncertain, developing an approach to controller design is a challenging task and has received significant attention from the research community. 
Currently, the choice and design of an appropriate controller for mobile robots is primarily driven by the intended application.

Parking/posture control is considered to be a representative problems in synthesizing controllers for mobile robots~\cite{tzafestas2013introduction}.  
The complex environments in which mobile robots operate motivates the use of intelligent control techniques.
In many parking control problems, Fuzzy Logic Control (FLC) is used to incorporate human knowledge to control the system intuitively, especially if it is difficult to develop model-based controllers.
Chang \textit{et al.} \cite{chang2002design}, investigated the use of FLC to solve the parallel parking problem for WMRs.

With respect to the MAMR, the authors in~\cite{nikshi2016parking} used FLC to control the $y$ position and orientation of the robot and then used a proportional controller with saturation to control the $x$ position to fully park the robot. 
In~\cite{nikshi2016parking}, the driving force was kept constant to simplify the problem.
The major limitation of using FLC in this application is the substantial increase of the fuzzy rule bases if the $x$ and $y$ positions were controlled simultaneously while operating under variable driving force.
In addition, for larger operating regions and more complicated tasks, constructing the rule bases can be difficult. 
To solve these problems, the authors in~\cite{Walelign2017SMC} proposed the use of Sliding Mode Control (SMC).
With this controller, the $x$ and $y$ positions were controlled simultaneously under variable driving force.
In addition, the SMC is more robust and is a natural control strategy for MAMR due to the discontinuous nature of the dynamics.
The authors proved in \cite{Walelign2017SMC} that the SMC can be used to park the MAMR from a given initial configuration to a final configuration.

This work builds on our previous work on the parking control problem of the MAMR presented in~\cite{Walelign2017SMC,Walelign2018_SMC_FLC} by demonstrating experimental performance of a SMC-based parking controller.
The MAMR presented in~\cite{Walelign2017SMC}, with its ball-type caster brakes and ability to move in any direction, is treated as holonomic.
However, the robot presented in~\cite{Walelign2018_SMC_FLC} and discussed in this paper incorporates conventional wheels and carries their nonholonomic constraints with them.
In this work, we present the experimental validation of the SMC in application to the MAMR.
In addition, this work shows the dynamics of the robots presented in~\cite{Walelign2017SMC} and~\cite{Walelign2018_SMC_FLC} remain similar for both reaching and sliding phases of the SMC.

The remainder of the paper is organized as follows. 
In Section \ref{model}, the mathematical modeling of the MAMR is presented. 
Section \ref{control} presents controller design and discussion. 
The experimental setup for real-time implementation is presented in Section \ref{experimental_setup}. 
Section \ref{experimental_result} gives the experimental results for parking the MAMR using SMC. 
Conclusions and directions for future work are presented in Section \ref{conclusion_recommendation}.
\begin{figure}[!t]
\centering
\includegraphics[width=\columnwidth ]{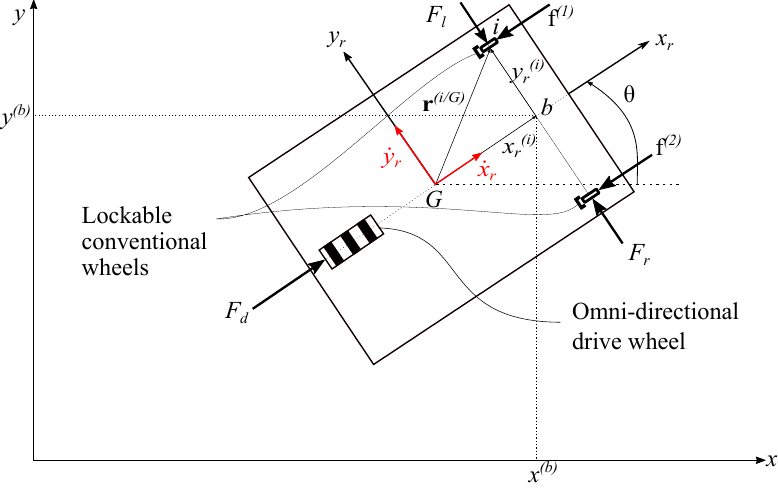}
\caption{Coordinate systems used for derivation of the equations of motion~\cite{Walelign2018_SMC_FLC}. }
\label{CoordinateSystems_MAMR2}
\end{figure}

\section{Mathematical Model} \label{model}

The equations of motion for the MAMR are derived in~\cite{Walelign2018_SMC_FLC} and revisited here for the sake of completeness.
Because of the nonholonomic constraints due to conventional wheels, the robot has less controllable degrees of freedom than the total degrees of freedom of the system.
The free body diagram of the robot with two coordinate systems, the global and local coordinate frames, is shown in Fig.~\ref{CoordinateSystems_MAMR2}. 
The global coordinate frame is denoted by ($x,y$) and the local coordinate frame by ($x_r, y_r$).
They also define the positions of the robot's center of mass in their respective coordinate frames.
The orientation of the robot is defined by $\theta$, which is the angle between the $x_r$-axis and $x$-axis.

The equations of motion can be derived in either of the coordinate frames and transformed from one to the other.
The details on the derivation of the equations of motion in the local coordinate frame can be found in~\cite{Walelign2018_SMC_FLC} and are summarized here as
%
%
\begin{align} 
\ddot{x}_r =&\frac{F_d}{m} - \sum_{i=1}^{2} \frac{g\mu^{(i)}_k  F^{(i)} }{3||\textbf{v}^{(i)}||} \left(\dot{x}_r - y^{(i)}_r \dot{\theta}\right) + \dot{y}_r \dot{\theta}, \label{2BareActiveDynamics_II_x} \\  
\ddot{y}_r  =& - \alpha\ddot{\theta},  \label{2BareActiveDynamics_II_y}  \\ 
\ddot{\theta}  =& \frac{1}{I + m{\alpha}^2} \sum_{i=1}^{2} \frac{mg \mu^{(i)}_k  F^{(i)} }{3||\textbf{v}^{(i)}||} \left( y^{(i)}_r \dot{x}_r  - {\left( y^{(i)}_r\right)}^2 \dot{\theta} \right) \nonumber \\ 
&  + \frac{m\alpha\dot{x}_r\dot{\theta}}{I + m{\alpha}^2}, \label{2BareActiveDynamics_II_z}
\end{align}
where $m$ is the total mass of the robot, $\mu^{(i)}_k$ is the kinetic coefficient of friction, $F_d$ is the driving force, $F^{(i)} \in \{0, 1\}$ is a discrete state that describes the state of brake $i$, $I$ is the moment of inertia about the center of mass, $x^{(i)}_r$ and $y^{(i)}_r$ are the $x_r$ and $y_r$ components of brake $i$, respectively.
$\alpha$ replaced $x^{(1)}_r$ for terms that are outside the summation signs in (\ref{2BareActiveDynamics_II_y}) and (\ref{2BareActiveDynamics_II_z}) to avoid confusion with the notation.
$\dot{x}_r$ is the velocity along the $x_r$-axis, $\dot{y}_r$ is the velocity along the $y_r$-axis, $\dot{\theta}$ is the angular velocity, $\ddot{x}_r$ is the acceleration along the $x_r$-axis, $\ddot{y}_r$ is the acceleration along the $y_r$-axis, and $\ddot{\theta}$ is the angular acceleration of the robot. 
$\textbf{v}^{(i)}$ is the velocity vector of brake $i$ given by
\begin{equation}
\textbf{v}^{(i)} = \begin{bmatrix}
             \ \dot{x}_r - y^{(i)}_r \dot{\theta}  \\
             \  0 \\
             \end{bmatrix},
\label{BrakeVelocity_2}     
\end{equation}  
in the local coordinate frame and $||\textbf{v}^{(i)}||$ is the magnitude of the velocity vector which is given by 
\begin{equation}
||\textbf{v}^{(i)}|| = |\dot{x}_r - y^{(i)}_r\dot{\theta} |.
 \label{velocity_Vector_II}
\end{equation}

In this work, the analysis is done in the global coordinate frame for the sake of simplicity.
The local equations of motion are transformed to the global coordinate frame using the orthogonal transformation matrix, $\textbf{R}(\theta)$, given by
\begin{equation}
\textbf{R}(\theta) = \begin{bmatrix}
\cos\theta&-\sin\theta&0\\
\sin\theta&\cos\theta&0\\
0&0&1\\
\end{bmatrix}.
\label{RotationMatrix} 
\end{equation}
The local and global motions of the robot are related by
\begin{equation}
\textbf{q} = {\textbf{R}(\theta)}{\textbf{q}_r},
\label{CoordinateTransformation} 
\end{equation} 
where $\textbf{q}_r$ and $\textbf{q}$ can be the local and global positions, velocities, or accelerations, respectively.
It is important to note that the position, velocity, and acceleration needs to be defined properly in each coordinate frame. 
The positions are defined as $\textbf{q}_r=[x_r,\: y_r,\: \theta]^T$ and $\textbf{q}=[x,\:y,\:\theta]^T$, the velocities as $\dot{\textbf{q}}_r=[\dot{x}_r,\:\dot{y}_r,\: \dot{\theta}]^T$  and $\dot{\textbf{q}}=[\dot{x},\:\dot{y},\: \dot{\theta}]^T$, and the accelerations as $\ddot{\textbf{q}}_r=[\ddot{x}_r - \dot{y}_r \dot{\theta},\,\,\,\, \ddot{y}_r + \dot{x}_r \dot{\theta},\: \ddot{\theta}]^T$ and $\ddot{\textbf{q}}=[\ddot{x},\: \ddot{y},\: \ddot{\theta}]^T$ in the local and global coordinate frames, respectively.
\begin{figure}[!t]
\centering
\includegraphics[width=\columnwidth ]{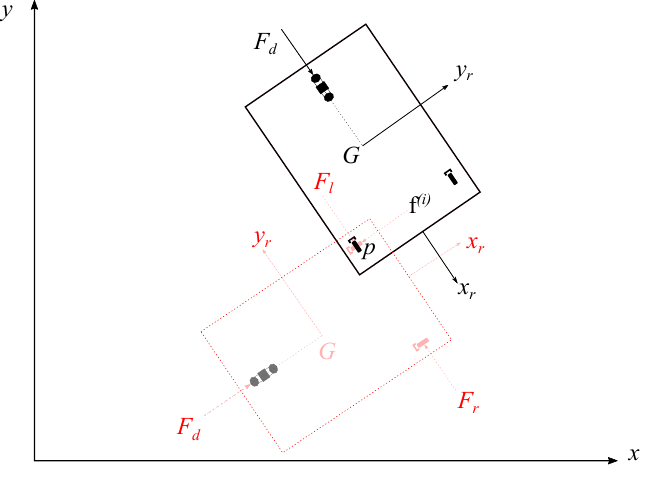}
\caption{The free body diagram of the MAMR under fixed axis rotation approximation.}
\label{CoordinateSystems_MAMR2}
\end{figure}   

Using~(\ref{RotationMatrix}) and~(\ref{CoordinateTransformation}), the local equations of motion given by ~(\ref{2BareActiveDynamics_II_x}),~(\ref{2BareActiveDynamics_II_y}), and~(\ref{2BareActiveDynamics_II_z}) are transformed to the equations of motion in the global coordinate frame as
\begin{align} 
\ddot{x}  =& \frac{F_d\cos\theta}{m}  -  \sum_{i=1}^{2} \frac{g\mu^{(i)}_k  F^{(i)}}{3||\textbf{v}^{(i)}||} \left( \dot{x}^{(i)}  \right) +  \alpha \ddot{\theta} \sin\theta  \nonumber \\ 
      & - \dot{\theta}\sin\theta\left(\dot{x} \cos\theta + \dot{y} \sin\theta\right), \label{MAMR2_dynamics_x} \\
\ddot{y} =& \frac{F_d\sin\theta}{m}  - \sum_{i=1}^{2} \frac{g\mu^{(i)}_k  F^{(i)}}{3||\textbf{v}^{(i)}||} \left( \dot{y}^{(i)}  \right) -  \alpha \ddot{\theta} \cos\theta   \nonumber\\ 
      & + \dot{\theta }\cos\theta\left(\dot{x} \cos\theta + \dot{y} \sin\theta\right), \label{MAMR2_dynamics_y}\\
\ddot{\theta} =& -\frac{1}{I + m{\alpha}^2} \sum_{i=1}^{2} \frac{mg\mu^{(i)}_k  F^{(i)} }{3||\textbf{v}^{(i)}||} \left( x^{(i)} \dot{y}^{(i)}   -y^{(i)} \dot{x}^{(i)} \right)  \nonumber \\ 
      & + \frac{m\alpha\dot{\theta}\left(\dot{x} \cos\theta + \dot{y} \sin\theta\right)}{I + m{\alpha}^2}, \label{MAMR2_dynamics_z}
\end{align}
where $\dot{x}$ and $\dot{y}$ are the $x$ and $y$ velocities of the robot's center of mass in the global coordinate frame, respectively. 
$x^{(i)}$ and $y^{(i)}$ are the components of the position vector, $\textbf{r}^{(i/G)}$, in the global coordinate frame and are given by 
\begin{equation}
\textbf{r}^{(i/G)} = \begin{bmatrix}
    \  x^{(i)}\\
    \  y^{(i)}\\       
    \end{bmatrix} =  \begin{bmatrix}
    \  x^{(i)}_r\cos\theta - y^{(i)}_r\sin\theta\\
    \  x^{(i)}_r\sin\theta + y^{(i)}_r\cos\theta\\
    \end{bmatrix}.
\label{position_vector_global}
\end{equation}
The velocity vector of the braking point, $\textbf{v}^{(i)}$, in this case is decomposed in the global coordinate frame as $\dot{x}^{(i)}$ and $\dot{y}^{(i)}$, which are given by   
\begin{equation}
\textbf{v}^{(i)} = \begin{bmatrix}
   \ \dot{x}^{(i)}\\
   \  \dot{y}^{(i)}\\
   \end{bmatrix} = \begin{bmatrix}
   \ \dot{x} \cos^2\theta + \dot{y} \cos\theta \sin\theta - y^{(i)}_r  \dot{\theta} \cos\theta\\
   \  \dot{x} \cos\theta\sin\theta + \dot{y} \sin^2\theta - y^{(i)}_r  \dot{\theta} \sin\theta \\
   \end{bmatrix},
\label{velocity_i}     
\end{equation} 
and its magnitude, $||\textbf{v}^{(i)}||$, is given by 
\begin{equation}
||\textbf{v}^{(i)}|| = \sqrt{ {(\dot{x}^{(i)} )}^2 + {(\dot{y}^{(i)} )}^2}.
 \label{velocity_mag}
\end{equation} 

In a special case of~(\ref{MAMR2_dynamics_x}),~(\ref{MAMR2_dynamics_y}), and~(\ref{MAMR2_dynamics_z}), the dynamics of the MAMR can be treated as if the system were in fixed axis rotation for the case that there is only one brake active and that both the velocity and driving force are sufficiently small.
This makes intuitive sense given the assumption of a Coulomb friction model, as the effect of static friction will not be overcome for a sufficiently small force.
As a result, if one brake is locked, then the MAMR, driven by the omni-directional drive wheel, will pivot about this locked braking point.
The free body diagram of the MAMR for fixed axis rotation is shown in Fig.~\ref{CoordinateSystems_MAMR2}.
The equations of motion under this assumption are given by 
\begin{align}
& \ddot{x} = {\ddot{\theta}}^{(p)} y^{(p)}  + \left({\dot{\theta}}^{(p)}\right)^2 x^{(p)}, \label{Pendulum_dynamics_x}\\
& \ddot{y} = -{\ddot{\theta}}^{(p)} y^{(p)}  - \left(\dot{\theta}^{(p)}\right)^2 x^{(p)}, \label{Pendulum_dynamics_y}\\
& \ddot{\theta}^{(p)} = \frac{y^{(p)}_r}{I^{(p)}} {F_d}, \label{Pendulum_dynamics_t}
\end{align}
where the subscript $p$ indicates the active brake.
$I^{(p)}$ is the moment of inertia about the active braking point $p$ and is given by $I^{(p)} = I + m\left({\left( x^{(p)}_r\right)}^2 + {\left( y^{(p)}_r\right)}^2\right)$. 
$x^{(p)}_r$ and $y^{(p)}_r$ are the $x_r$ and $y_r$ components of the position 
of active brake $p$ in the local coordinate frame, respectively. 
\begin{figure}[!t]
\centering
\includegraphics[scale=1.0]{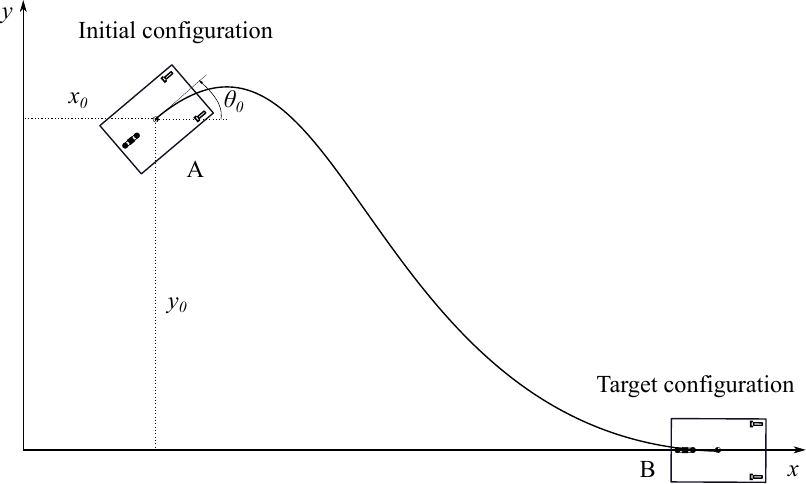} 
\caption{Parking control problem formulation for the MAMR~\cite{Walelign2018_SMC_FLC}. }
\label{ParkingProblem_MAMR2}
\end{figure}

\section{Control Design} \label{control}

In contrast to controlling the $x$ and $y$ positions of the MAMR sequentially using FLC as in~\cite{nikshi2016parking}, the objective here is to control the $x$ and $y$ positions simultaneously. The parking problem under consideration is illustrated in Fig.~\ref{ParkingProblem_MAMR2}.
The goal is to drive the MAMR from a given initial configuration $\mathcal{C}_0 = [x_0,y_0,\theta_0] \in SE(2)$ to a desired final configuration $\mathcal{C}_f = [x_f,y_f,\theta_f] \in SE(2)$ in finite time.
  

SMC can be seen as an obvious choice for the parking control of the MAMR because of the discontinuous dynamics imposed by the two-state braking actuators.
It also avoids the substantially increased complexity of the fuzzy rule bases that would be seen with the FLC of \cite{nikshi2016parking} if the $x$ and $y$ positions of the robot were controlled simultaneously while operating under variable driving force.
The design of the SMC with application to the MAMR was first introduced in~\cite{Walelign2017SMC} and is summarized here for the sake of completeness.
It is important to point out that the SMC designed in~\cite{Walelign2017SMC} was for the holonomic MAMR that used a ball-type caster as a brake.
This work, however, designs and implements the SMC to the nonholonomic MAMR that uses braked, free spinning conventional wheels.

\begin{figure}[!t]
\centering
\includegraphics[scale = 1.0]{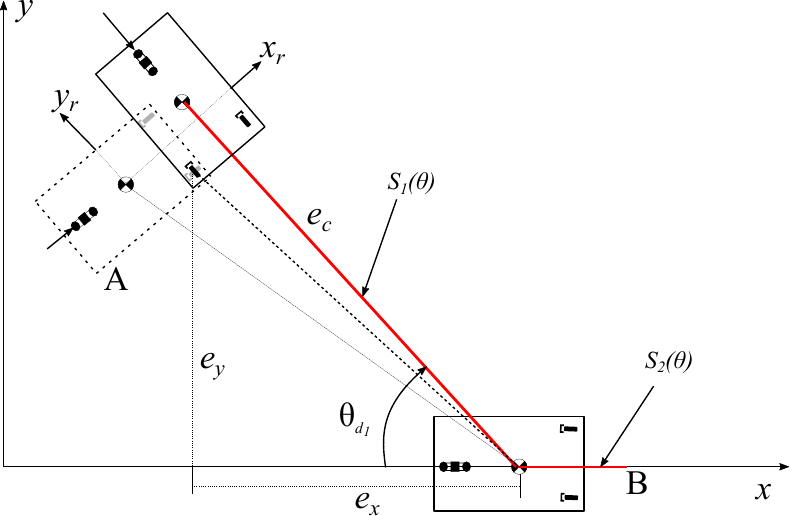} 
\caption{Sliding surfaces used for parking the MAMR from a given initial configuration to desired final configuration. }
\label{Slidingsurface_Jour}
\end{figure}

To park the MAMR at a desired final configuration from a given initial configuration, two sliding surfaces have been designed under the SMC framework. 
These are depicted in Fig.~\ref{Slidingsurface_Jour} where the first sliding surface, $S_1 (\theta)$, brings the robot closer to the target point and the second sliding surface, $S_2 (\theta)$, stabilizes the robot about its final configuration.
The robot is then brought to a stop while sliding on $S_2 (\theta)$ once the target point is attained.
The first sliding surface is defined as 
\begin{equation}
S_1(\theta) = \dot{e}_{\theta_1} + {\lambda}_1 e_{\theta_1},
\label{slidingDurface_1}
\end{equation}
where $e_{\theta_1} = \theta_{d_1} - \theta$ is the error in orientation and ${\lambda}_1$ is an arbitrary positive constant.
For this surface, the desired orientation is defined as 
\begin{equation}
\theta_{d_1} = \tan^{-1}\left(\frac{y_f - y}{x_f - x}\right),
\label{slidingDurface_angle}
\end{equation}
where $x_f$ and $y_f$ are the desired final $x$ and $y$ positions of the robot, respectively.
The second sliding surface has a similar form to $S_1 (\theta)$ and is defined as
\begin{equation}
S_2 (\theta) = \dot{e}_{\theta_1} + {\lambda}_2 e_{\theta_2},
\label{slidingDurface_2}
\end{equation}
where $e_{\theta_2} = \theta_{d_2} - \theta$ is, similarly, the error in orientation and ${\lambda}_2$ is another arbitrary positive constant. 
For this surface, however, the desired orientation is defined as $\theta_{d_2}  = 0$.
Note that any angle can be chosen as the final orientation.
From~(\ref{slidingDurface_1}) and~(\ref{slidingDurface_2}), one can see that if $S_1(\theta)$ and $S_2(\theta)$ converge to zero, $e_{\theta_1}$ and $e_{\theta_2}$ converge to zero trivially. 

Using these sliding surfaces, three control laws have been designed for the reaching and sliding phase of each sliding surface.
With the fixed axis rotation approximation made for the MAMR dynamics in~(\ref{Pendulum_dynamics_x}), (\ref{Pendulum_dynamics_y}), and (\ref{Pendulum_dynamics_t}), the reaching phase is simplified to a simple stabilizing controller about the desired angle.
The equations of motion given by~(\ref{MAMR2_dynamics_x}),~(\ref{MAMR2_dynamics_y}), and~(\ref{MAMR2_dynamics_z}) can then be used for the design of the sliding part of the control law.
For the stopping phase, the dynamics of the MAMR are approximated as a one-dimensional mass system with unknown coefficient of friction, but bounded by $[0 \,\,\,\,1]$, under the assumption of sufficiently fast switching of the brakes.
These control laws in application to parking the nonholonomic MAMR from a given initial configuration to a desired final configuration is proved here. 

Therefore, using~(\ref{Pendulum_dynamics_x}), (\ref{Pendulum_dynamics_y}), and (\ref{Pendulum_dynamics_t}) for the reaching
phase and~(\ref{MAMR2_dynamics_x}),~(\ref{MAMR2_dynamics_y}), and~(\ref{MAMR2_dynamics_z}) for the sliding phase; three
controllers are designed for each sliding surface: controller for the reaching phase, the
sliding phase, and the stopping phase. 
By switching among the two sliding surfaces and the three controllers, the MAMR is driven from any initial configuration $\mathcal{C}_0$ to a desired final configuration $\mathcal{C}_f$ in finite time. 
Without loss of generality, the three controllers are designed for the first sliding surface $S_1(\theta)$. These controllers include:
\begin{enumerate}
\item \underline{Reaching phase:} Regardless of the robot's initial orientation in the $x-y$  plane, the robot is made to orient about the desired sliding angle $\theta_{d_1}$ to reach the sliding surface.
To achieve this orientation, the controller given by $u = [F_d, \: 1, \: 0]^T$ when Brake $1$ is activated or $u = [F_d, \: 0, \: 1]^T$ when Brake $2$ is activated is proposed.
Note that $F^{(i)}$ is set to $1$ for the active brake and $0$ for the inactive brake.
The driving force $F_d$ is the stabilizing control law given by 
\begin{equation}
F_d = K_p{e_{\theta_1}} + K_d \dot{e}_{\theta_1},
\label{drivingForce}
\end{equation}
where $K_p$ and $K_d$ are the proportional and derivative gains, respectively.
In this case, the equations for fixed axis rotation given by~(\ref{Pendulum_dynamics_x}), (\ref{Pendulum_dynamics_y}), and (\ref{Pendulum_dynamics_t}) are used.
The desired angle ($\theta_{d_1}$) is assumed to be constant for the sake of simplicity in analyses.
With this assumption, the second time derivative of the orientation error ($e_{\theta_1}$) is found as $\ddot{e}_{\theta_1} =  -{\ddot{\theta}}$. 
Using the controller for the driving force given by Equation (\ref{drivingForce}), the angular acceleration $\ddot{\theta} = -\ddot{e}_{\theta_1}$, and replacing $\frac{y^{(p)}_r}{I^{(p)}}$ by a constant $c$ in Equation~(\ref{Pendulum_dynamics_t}) results in
\begin{equation}
\ddot{e}_{\theta_1} + cK_d \dot{e}_{\theta_1} + cK_pe_{\theta_1}  = 0.
\label{errorDyna}
\end{equation}
From the error dynamics given by~(\ref{errorDyna}), it can be shown that $e_{\theta_1} \rightarrow 0$ as $t \rightarrow \infty$ for $K_p > 0$ and $K_d > 0$. 
Furthermore, the system will be overdamped if the gain $K_d$ is chosen as $K_d \geq \sqrt{4K_p/c}$.  
 
\item \underline{Sliding phase:} 
As opposed to the Ackermann steered mobile robots, the MAMR will not slide on an arbitrary sliding surface. 
The curvature of the sliding surface for the MAMR is limited by the robot geometry. 
For example, for high curvature the brakes will not have enough control authority to keep the robot on the surface when moving at high speed. 
However, for a straight line, once on the sliding surface, it will remain moving on the 
sliding surface for all driving force under sufficiently fast switching of the two brakes.
Toward this end, controllers given by $u = [F_d, \: F^{(1)}, \: F^{(2)}]^T$ are used to achieve the objective, where $F^{(1)} = \frac{1}{2} \left(1+\text{sgn}(S_1(\theta))\right)$, $F^{(2)} = \frac{1}{2}\left(1-\text{sgn}(S_1(\theta))\right)$, and $F_d$ is any driving force to drive the robot along the sliding surface. 
To prove this, the same procedure as~\cite{Walelign2017SMC} is followed except the equations of motion given by~(\ref{MAMR2_dynamics_x}),~(\ref{MAMR2_dynamics_y}), and~(\ref{MAMR2_dynamics_z}) are used. 
The necessary and sufficient conditions for the trajectories to stay on the sliding surface are $S_1(\theta) = 0$ and $\dot{S}_1(\theta) = 0$ \cite{sira1987variable}. 
In addition, the following conditions are trivially true on the sliding surface  
\begin{enumerate}
\item The two brakes are off, $\mu^{(i)}_k = 0$ where $i \in \{1, 2\}$.
\item The orientation error, $e_{\theta_1}$, is zero, resulting in $\theta = \theta_{d_1}$.
\item The time derivative of the orientation error, $\dot{e}_{\theta_1}$, is zero, i.e., $\dot{\theta}  = \dot{\theta}_{d_1} = 0$.
\end{enumerate}

Differentiating~(\ref{slidingDurface_1}) with respect to time, using the relation for the error along the sliding surface, $e_c$, as
\begin{equation}
 e_c  = \sqrt{e_x^2 + e_y^2},
\label{fig:geometricRelationEc}
\end{equation}
with the errors in the $x$ and $y$ positions being defined as 
\begin{equation}
\begin{aligned}
 & e_x = x_f - x = -e_c \cos{\theta},\\
 & e_y = y_f - y = -e_c \sin{\theta},\\
\end{aligned}
\label{fig:geometricRelation}
\end{equation}
together with using the conditions listed above for the sliding surface, and replacing $\theta$ by $\theta_{d_1}$, the sliding surface dynamics becomes
\begin{equation} \label{slidingDynamics_1}
\begin{aligned} 
&\resizebox{0.8\columnwidth}{!}{$ \dot{S}_1  = - {\dot{y}}\, \left(-\frac{2\, {{e_y}}^2\, {\dot{x}}}{{{e_x}}^4\, {\left(\frac{{{e_y}}^2}{{{e_x}}^2} + 1\right)}^2} + \frac{{\dot{x}}}{{{e_x}}^2\, \left(\frac{{{e_y}}^2}{{{e_x}}^2} + 1\right)} + \frac{2\, {e_y}\, {\dot{y}}}{{{e_x}}^3\, {\left(\frac{{{e_y}}^2}{{{e_x}}^2} + 1\right)}^2}\right) $} \\ 
 & \resizebox{0.8\columnwidth}{!}{$ -{\dot{x}}\, \left(\frac{{\dot{y}}}{{{e_x}}^2\, \left(\frac{{{e_y}}^2}{{{e_x}}^2} + 1\right)} + \frac{2\, {{e_y}}^3\, {\dot{x}}}{{{e_x}}^5\, {\left(\frac{{{e_y}}^2}{{{e_x}}^2} + 1\right)}^2} - \frac{2\, {{e_y}}^2\, {\dot{y}}}{{{e_x}}^4\, {\left(\frac{{{e_y}}^2}{{{e_x}}^2} + 1\right)}^2} - \frac{2\, {e_y}\, {\dot{x}}}{{{e_x}}^3\, \left(\frac{{{e_y}}^2}{{{e_x}}^2} + 1\right)}\right) $} \\
 & \resizebox{0.8\columnwidth}{!}{$+ \left( \frac{e_y \frac{F_d \cos\theta_{d_1}}{m} - e_x \frac{F_d \sin\theta_{d_1}}{m}}{{e_x}^2 \left(\frac{{e_y}^2}{{e_x}^2}+1\right)} - \frac{m {\alpha}^2 \dot{\theta}_{d_1}  \left(\dot{x}\cos\theta_{d_1} + \dot{y}\sin\theta_{d_1}\right)}{{e_x}^2 \left(\frac{{e_y}^2}{{e_x}^2}+1\right)} \right) $} \\  
 & \resizebox{0.9\columnwidth}{!}{$ + \frac{e_y \left( -\dot{\theta}_{d_1}\sin\theta_{d_1}\left(\dot{x}\cos\theta_{d_1} + \dot{y}\sin\theta_{d_1}\right)  + \frac{m {\alpha}^2 \dot{\theta}_{d_1} \sin\theta_{d_1} \left(\dot{x}\cos\theta_{d_1} + \dot{y}\sin\theta_{d_1}\right)}{m {\alpha}^2 +I} \right)}{{e_x}^2 \left(\frac{{e_y}^2}{{e_x}^2}+1\right)}  $} \\  
 & \resizebox{0.9\columnwidth}{!}{$  +\frac{e_x \left( - \dot{\theta}_{d_1}\cos\theta_{d_1}\left(\dot{x}\cos\theta_{d_1} + \dot{y}\sin\theta_{d_1}\right)  + \frac{m {\alpha}^2 \dot{\theta}_{d_1} \cos\theta_{d_1} \left(\dot{x}\cos\theta_{d_1} + \dot{y}\sin\theta_{d_1}\right)}{m {\alpha}^2 +I} \right)}{{e_x}^2 \left(\frac{{e_y}^2}{{e_x}^2}+1\right)}  $} \\ 
\end{aligned}.
\end{equation}
Because the derivative of the desired angle on the sliding surface is zero (i.e., $\dot{\theta}_{d_1} = 0$), all terms that are multiple of $\dot{\theta}_{d_1}$ in~(\ref{slidingDynamics_1}) vanishes and results in 
\begin{equation} \label{slidingDynamics_1_reduce}
\begin{aligned} 
&\resizebox{0.9\columnwidth}{!}{$ \dot{S}_1  = - {\dot{y}}\, \left(-\frac{2\, {{e_y}}^2\, {\dot{x}}}{{{e_x}}^4\, {\left(\frac{{{e_y}}^2}{{{e_x}}^2} + 1\right)}^2} + \frac{{\dot{x}}}{{{e_x}}^2\, \left(\frac{{{e_y}}^2}{{{e_x}}^2} + 1\right)} + \frac{2\, {e_y}\, {\dot{y}}}{{{e_x}}^3\, {\left(\frac{{{e_y}}^2}{{{e_x}}^2} + 1\right)}^2}\right) $} \\ 
 & \resizebox{0.9\columnwidth}{!}{$ -{\dot{x}}\, \left(\frac{{\dot{y}}}{{{e_x}}^2\, \left(\frac{{{e_y}}^2}{{{e_x}}^2} + 1\right)} + \frac{2\, {{e_y}}^3\, {\dot{x}}}{{{e_x}}^5\, {\left(\frac{{{e_y}}^2}{{{e_x}}^2} + 1\right)}^2} - \frac{2\, {{e_y}}^2\, {\dot{y}}}{{{e_x}}^4\, {\left(\frac{{{e_y}}^2}{{{e_x}}^2} + 1\right)}^2} - \frac{2\, {e_y}\, {\dot{x}}}{{{e_x}}^3\, \left(\frac{{{e_y}}^2}{{{e_x}}^2} + 1\right)}\right) $} \\
 & \resizebox{0.6\columnwidth}{!}{$ -\frac{F_d\, \sin\!\theta_{d_1}}{{e_x}\, m\, \left(\frac{{{e_y}}^2}{{{e_x}}^2} + 1\right)} + \frac{F_d\, {e_y}\, \cos\!\theta_{d_1}}{{{e_x}}^2\, m\, \left(\frac{{{e_y}}^2}{{{e_x}}^2} + 1\right)} = 0. $}  
\end{aligned}
\end{equation}
Equation~(\ref{slidingDynamics_1_reduce}) is exactly the same as the equation for the sliding surface dynamics for the holonomic MAMR presented in~\cite{Walelign2017SMC} and hence the same procedure is followed. 

Assume that $e_x \neq 0$ to avoid division by zero.
Using the relation between the error in positions and the desired angle as $\frac{y_f - y}{x_f - x}= \frac{e_y}{e_x} = \tan\theta_{d_1}$, and dividing~(\ref{slidingDynamics_1_reduce}) by $e_x$, the coefficient of $F_d$ vanishes resulting in
\begin{equation} \label{fig:slidingDynamics_2}
\begin{split}
\dot{S}_1  & = \resizebox{0.8\hsize}{!}{$-{\dot{y}}\, \left(-\frac{2\, {{e_y}}^2\, {\dot{x}}}{{{e_x}}^5\, {\left(\frac{{{e_y}}^2}{{{e_x}}^2} + 1\right)}^2} + \frac{{\dot{x}}}{{{e_x}}^3\, \left(\frac{{{e_y}}^2}{{{e_x}}^2} + 1\right)} + \frac{2\, {e_y}\, {\dot{y}}}{{{e_x}}^4\, {\left(\frac{{{e_y}}^2}{{{e_x}}^2} + 1\right)}^2}\right)$} \\
 & \resizebox{.9\hsize}{!}{$-{\dot{x}}\, \left(\frac{{\dot{y}}}{{{e_x}}^3\, \left(\frac{{{e_y}}^2}{{{e_x}}^2} + 1\right)} + \frac{2\, {{e_y}}^3\, {\dot{x}}}{{{e_x}}^6\, {\left(\frac{{{e_y}}^2}{{{e_x}}^2} + 1\right)}^2} - \frac{2\, {{e_y}}^2\, {\dot{y}}}{{{e_x}}^5\, {\left(\frac{{{e_y}}^2}{{{e_x}}^2} + 1\right)}^2} - \frac{2\, {e_y}\, {\dot{x}}}{{{e_x}}^4\, \left(\frac{{{e_y}}^2}{{{e_x}}^2} + 1\right)}\right) = 0.$} \\
\end{split}
\end{equation}
Equation~(\ref{fig:slidingDynamics_2}) can be further simplified by dividing both sides by ${{{e_x}}^3\, \left(\frac{{{e_y}}^2}{{{e_x}}^2} + 1\right)}$ and substituting the time derivative of the position errors given by~(\ref{fig:geometricRelation}) yields
\begin{eqnarray} \label{fig:slidingDynamics_5}
\dot{S}_1  &=& {2\dot{x}\dot{y}}\left(1 - 2\cos^{2}\theta_{d_1}\right) + 2{\dot{x}}^2 \sin\theta_{d_1}\cos\theta_{d_1} \nonumber  \\ 
      & & - 2{\dot{y}}^2\sin\theta_{d_1}\cos\theta_{d_1}= 0. 
\end{eqnarray}
Using the relation for double angle formula, $1 - 2{\cos^{2}}\theta_{d_1} = -\cos2\theta_{d_1}$ and $2\sin\theta_{d_1}\cos\theta_{d_1} = \sin2\theta_{d_1}$,~(\ref{fig:slidingDynamics_5}) simplifies to
\begin{equation}
\frac{\sin2\theta_{d_1}}{\cos2\theta_{d_1}} = \frac{2\dot{x}\dot{y}}{{\dot{x}}^2 - {\dot{y}}^2}.
\label{fig:slidingDynamics_6}
\end{equation}
Writing the $x$ and $y$ velocities (i.e., $\dot{x}$ and $\dot{y}$) in terms of the time derivative of the error along the sliding surface ($\dot{e}_c$) by differentiating~(\ref{fig:geometricRelation}), and using the double angle formula again,~(\ref{fig:slidingDynamics_6}) reduces to
%
\begin{equation}
 \sin\left( 2{\theta}_{d_1}\right) \cos\left( 2{\theta}_{d_1} \right)  = \sin\left( 2{\theta}_{d_1} \right) \cos\left( 2{\theta}_{d_1} \right) ,
\label{fig:slidingDynamics_7}
\end{equation}
which proves $\dot{S}_1(\theta) = 0$ for any driving force ($F_d$) and hence the trajectory, once on the sliding surface, stays on the surface for all driving forces all the time under the assumption that the brakes will not saturate.
The same derivation also applies for the second sliding surface given by~(\ref{slidingDurface_2}).

\item \underline{Stopping phase:} 
It may be desirable to stop the MAMR while sliding along the sliding surface at some desired value, $e_c$.
From Fig. \ref{Slidingsurface_Jour}, one can see that the desired value is $e_c = 0$.
To stabilize the robot about this point, the control law $u = -K \textbf{x}$ is used, where $u = F_d$, the state vector $\textbf{x} = [e_c \,\,,\,\,\, \dot{e}_c]^T$, and the vector of control gains as $K = [K_1 \,\,\,\,\, K_2]$.
For this control scheme the system is treated as a one-dimensional mass system with Coulomb friction since it has been shown that the previous control law keeps the robot on the sliding surface.
Because of the fast switching of the two brakes, the coefficient of friction is unknown but bounded by $0$ from below and $1$ from above. 

To prove the stability of the system under the state feedback control law, the discontinuous Coulomb friction model is approximated by a continuous hyperbolic tangent function as 
\begin{equation}
F_f = \gamma_1\tanh(\gamma_2 \dot{e}_c),
\label{FrictionForce_continuous}
\end{equation}
where $\gamma_1 = \mu_k \frac{mg}{3}$ and $\gamma_2$ is a constant number. A large value of $\gamma_2$ is recommended for better approximation of the discontinuous friction model by hyperbolic tangent function. 
With this assumption, the error dynamics on the sliding surface is given by
\begin{equation}
\ddot{e}_c = \frac{F_d}{m} - \frac{\gamma_1}{m} \tanh(\gamma_2 \dot{e}_c).
\label{1Dequation}
\end{equation}
If the states are defined as $x_1 = e_c$ and  $x_2 = \dot{e}_c$, and the control input as $F_d= u$, the dynamics in state space form becomes
\begin{equation}
\dot{\textbf{x}} = \begin{bmatrix}
x_2\\
-\frac{\gamma_1}{m} \tanh(\gamma_2 x_2) + \frac{u}{m}\\
\end{bmatrix}. 
\label{stateSpaceModel} 
\end{equation} 
Since the goal is to stabilize the system about $(0,0)$, linearizing~(\ref{stateSpaceModel}) about this point yields 
\begin{equation}
 \dot{\textbf{x}} = \underbrace{\begin{bmatrix}
0 &1\\
0 &-\frac{\gamma_1 \gamma_2}{m} \\
\end{bmatrix} }_A \textbf{x} + \underbrace{\begin{bmatrix}
0\\
\frac{1}{m} \\
\end{bmatrix}}_B u.
\label{LinearstateSpaceModel} 
\end{equation} 
Because the pair $(A,B)$ in~(\ref{LinearstateSpaceModel}) is full rank, the system is controllable and hence a state feedback control law can be designed to stabilize the robot about $\textbf{x}  = 0$. 
The poles of the closed loop dynamics, $(A-BK)$, can be placed to any desired point using the control law given by $u = -K\textbf{x} $. 

Substituting the linear control law ($u = -K\textbf{x} $) into the nonlinear dynamics given by~(\ref{stateSpaceModel}), the closed loop dynamics becomes
\begin{equation}
\dot{\textbf{x}} = \begin{bmatrix}
x_2\\
-\frac{\gamma_1}{m} \tanh(\gamma_2 x_2) - \frac{K_1}{m}x_1 - \frac{K_2}{m}x_2\\
\end{bmatrix}. 
\label{NonlinearClosedLoop} 
\end{equation} 
The stability of the closed loop system given by~(\ref{NonlinearClosedLoop}) can be proved by using Lyapunov's direct method.
A candidate function given by
\begin{equation}
V(x) = \frac{1}{2} {\frac{K_1}{m}} {x_1}^2 + \frac{1}{2} {x_2}^2,
\label{LyapunovCandidate} 
\end{equation} 
with $K_1>0$ is considered. 
Differentiating the candidate function with respect to time results in
\begin{equation}
\dot{V}(x) = -\frac{\gamma_1}{m} x_2\tanh(\gamma_2 x_2) - \frac{K_2}{m} {x_2}^2 \leq 0.
\label{LyapunovCandidateDerivative} 
\end{equation}
With the coefficient of friction being bounded as $\mu_k \in [0, 1]$ and $\gamma_1 \geq 0$, the term $x_2\tanh(\gamma_2 x_2) \geq 0$ for $\forall x_2$, making the first term in~(\ref{LyapunovCandidateDerivative}) negative. 
For~(\ref{LyapunovCandidateDerivative}) to be negative semi-definite (i.e., $\dot{V}(x) \leq 0$), the gain $K_2$ has to be chosen as $K_2 > 0$.
Therefore, by choosing the gains as $K_1 > 0$ and $K_2 > 0$ for the linearized system, a sufficient condition for local stability is attained as $\dot{V}(x) \leq 0$, and hence the closed loop nonlinear system is locally stable at $\textbf{x}  = 0$.
In addition, because the Lyapunov candidate function given by~(\ref{LyapunovCandidate}) is radially unbounded, i.e., $V(x)\rightarrow \infty$ as $||x|| \rightarrow \infty$, the equilibrium point $\textbf{x}  = 0$ is globally stable.

In extension to the above results, the global asymptotic stability of the equilibrium point is checked by Lasalle's invariant principle.
If $\dot{V}(x)$ is identically zero over a nonzero time interval, the system trajectories can be computed by setting~(\ref{LyapunovCandidateDerivative}) to zero as
\begin{equation}
\dot{V}(x) = -\frac{\gamma_1}{m} x_2\tanh(\gamma_2 x_2) - \frac{K_2}{m} {x_2}^2 = 0.
\label{solvingforX2} 
\end{equation} 
Solving for $x_2$ from~(\ref{solvingforX2}) results in $x_2 = 0 $ $\forall t $ $\Longrightarrow$ $\dot{x}_2 = 0$. 
Then, substituting $x_2 = 0$ and $\dot{x}_2 = 0$ into~(\ref{NonlinearClosedLoop}) results in
 \begin{equation}
0 = -\frac{\gamma_1}{m} \tanh(\gamma_2 x_2) - \frac{K_1}{m}x_1 - \frac{K_2}{m}x_2\Longrightarrow x_1 = 0.
\label{solvingforX1} 
\end{equation} 
Therefore, $\dot{V}(x)$ does not vanish along the system trajectories other than $\textbf{x} =0$, and hence the equilibrium point of the closed loop system ($\textbf{x} =0$) is globally asymptotically stable. 
\end{enumerate}
%
\begin{figure}[t]
\centering
\includegraphics[width=\columnwidth ]{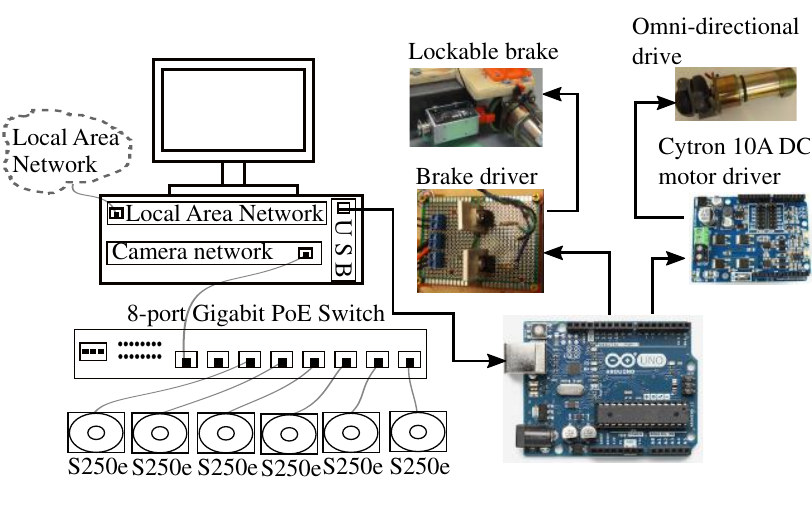} 
\caption{Workflow diagram for implementation of controllers in real-time. }
\label{Streamingflow_MAMR_II}
\end{figure}



\section{Experimental Setup} \label{experimental_setup}

The experimental setup used for validation of the controller designed in Section \ref{control} is presented here. 
The setup is similar to the one presented in~\cite{Walelign2018_SMC_FLC}.
The robot used for the experiments in this work uses a rubber wheel together with a solenoid as a brake to increase the coefficient of friction between the wheels and the ground.
As a result, when one brake is activated and for sufficiently small driving force, the robot will pivot about the active brake and the dynamics are approximated as fixed axis rotation, which was the assumption made for the reaching phase.  
In addition, the motor driver for the drive wheel was replaced by a Cytron $10A$ DC motor driver Arduino shield to improve driving force.

The position and orientation information of the robot are obtained using an OptiTrack camera system.
The cameras are mounted high on the wall to track the MAMR in the capture volume. 
The capture volume is constructed in such a way that at least three cameras are able to see the robot at a given time. 
Three spherical reflective markers are mounted on the robot in an asymmetric manner for tracking reliability and for the software to identify the orientation properly. 

The overall work flow for the experiment is shown in Fig.~\ref{Streamingflow_MAMR_II}. 
The cameras are connected to the host computer through an Ethernet connection via an 8-port Gigabit Power over Ethernet (PoE) switch.
The position and orientation of the MAMR are streamed to \textsc{Matlab}\textsuperscript{\textregistered}~\footnote{MATLAB and SIMULINK are registered trademarks of MathWorks, Inc., Natick, MA,USA.} using the IP address for the network in real-time.  
Then, the control signals are computed and the \textsc{Matlab}-Arduino communication package is used to send the control signal to the MAMR. 
The support package enables \textsc{Matlab} to communicate with an Arduino board over a USB cable from \textsc{Matlab}'s command line. The package is based on a server program running on the Arduino board, which listens to commands arriving via the serial port, executes the commands, and returns a result if needed.
In this experiment, the tracking software is run in the background with appropriate settings that allow real-time streaming to \textsc{Matlab}.
\begin{figure}[t]
\centering
\includegraphics[width=\columnwidth]{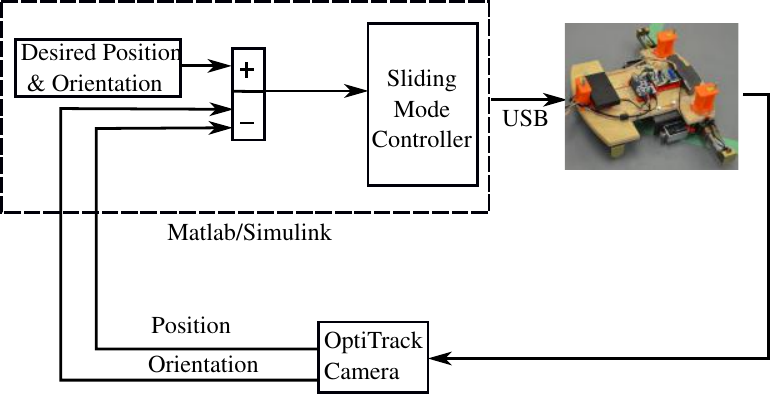}
\caption{Block diagram for hardware control implementation. }
\label{ClosedLoop_MAMR_II}
\end{figure}


\section{Experimental Results} \label{experimental_result}

In this section, the effectiveness of the controller designed in Section~\ref{control} is validated through experiments.
The block diagram for hardware control implementation is shown in Fig.~\ref{ClosedLoop_MAMR_II}.
The goal is to park the MAMR from any initial configuration $\mathcal{C}_0 = [x_0, y_0, \theta_0]$ to a desired final configuration $\mathcal{C}_f = [x_f, y_f, \theta_f]$ in some finite time.
To park the robot at the target configuration, we design the sliding surfaces given by~(\ref{slidingDurface_1}) and~(\ref{slidingDurface_2}) from the initial configuration, and then switch between the reaching and sliding modes of the SMC.
Note that some initial configurations may not need the two sliding surfaces to achieve the objective.
For example, if the robot is aligned with the $x$-axis at the initial point, it only uses $S_2(\theta)$ to park the robot at the final configuration along the same axis. 

\begin{figure}[t]
\centering
\includegraphics[width=\columnwidth ]{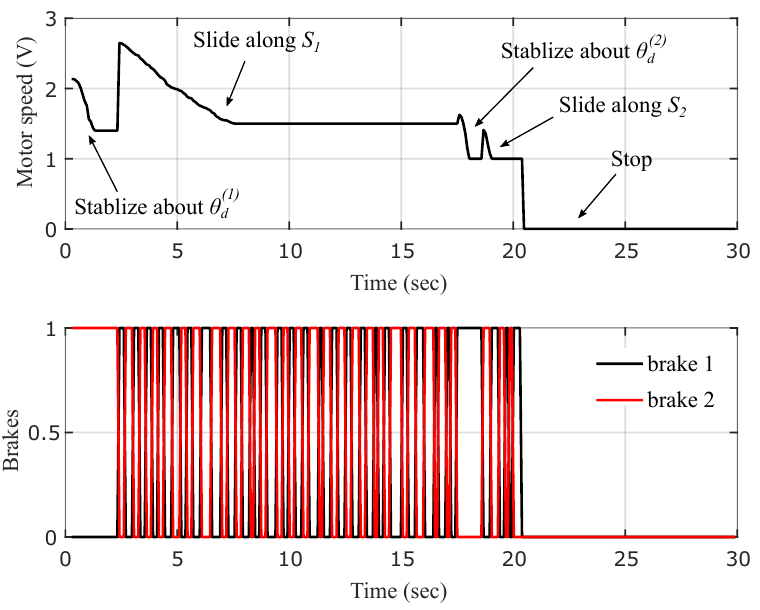} 
\caption{The plot for the driving force and the two brakes for parking the MAMR at final configuration of $\mathcal{C}_f = [2.00, 0.00, 0.00]$.}
\label{parking_control_effort}
\end{figure}

\begin{figure}[t]
\centering
\includegraphics[width=\columnwidth ]{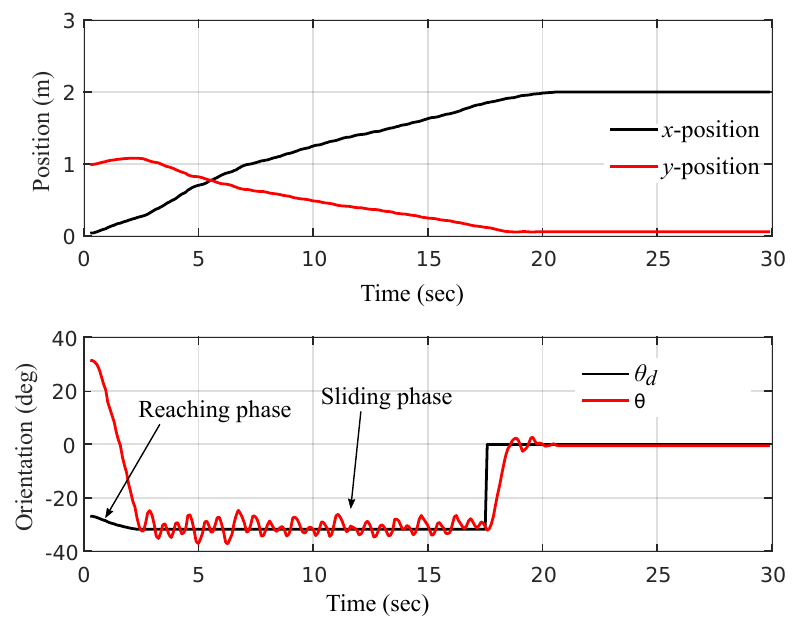} 
\caption{Position and orientation plot for parking the MAMR at final configuration of $\mathcal{C}_f = [2.00, 0.00, 0.00]$.}
\label{parking_control_configuration}
\end{figure}

The experimental results for parking the MAMR at point $\text{B}$ from point $\text{A}$ are shown in Fig.~\ref{parking_control_effort} and Fig.~\ref{parking_control_configuration}.
The robot's initial configuration is defined as $\mathcal{C}_0 = [0.00, 1.00, 30.00]$ and the desired final configuration is defined as $\mathcal{C}_f = [2.00, 0.00, 0.00]$, where we remind the reader that the units on $\mathbb{R}^2$ are in meters and the units on $\mathbb{S}^1$ are in degrees.
The control effort (i.e., continuous driving force) and the status of the two brakes (i.e., discrete ON or OFF) are shown in Fig.~\ref{parking_control_effort}. 
While it is possible to activate any of the brakes at the beginning, we chose to activate the brake closer to the target point.
In this case, Brake $2$ was activated at initial point to bring the robot to the sliding surface.
Once on the sliding surface ($S_1(\theta)$), the brakes switch between Brake $1$ and $2$ to keep the robot on the sliding surface until it gets closer to the target point.
When the robot is close to the target point, the controller switches to the reaching mode by activating Brake $1$ to make the robot orientation align with the target orientation.
Finally, the robot drives along the second sliding surface ($S_2(\theta)$) by minimizing the error in the $x$ direction until the target point is attained.
The plot for the driving force has four parts as shown in Fig.~\ref{parking_control_effort} (top): reaching $S_1(\theta)$, driving the robot along $S_1(\theta)$, reaching $S_2(\theta)$, and driving the robot along $S_2(\theta)$.
For both reaching and sliding modes, the driving force was bounded from below because of a deadzone present in the drive motor. 
It is also important to point out that the sliding surface is time varying during the reaching phase but remains constant during the sliding phase as shown in Fig.~\ref{parking_control_configuration} (bottom).


\begin{figure}[t]
\centering
\includegraphics[width=\columnwidth ]{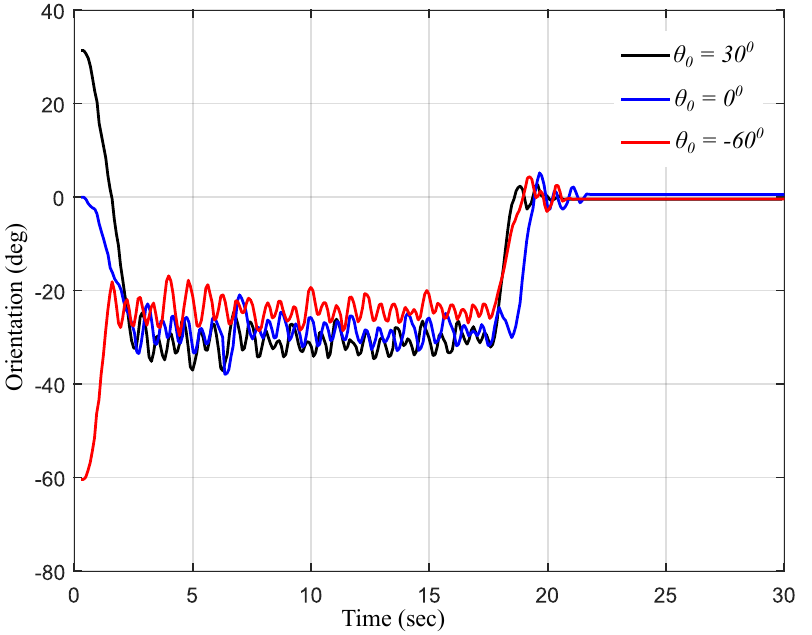} 
\caption{The plot for orientation for parking the robot from $\mathcal{C}_0=[0.00,1.00,\theta_0]$ with $\theta_0$ given by $\theta_0 = 30^\circ$, $\theta_0 = 0^\circ$, and $\theta_0 = -60^\circ$ to the same final configuration of $\mathcal{C}_f=[2.00,0.00,0.00]$. 
}
\label{orientation_different_IC}
\end{figure}

In Section \ref{control}, it was shown that the robot will converge to the desired final configuration regardless of its initial configuration. 
To show this, different experiments were done for some representative initial configurations.
Without loss of generality, the $x$ and $y$ positions were kept the same but different values for orientation were considered.
%
Fig.~\ref{orientation_different_IC} shows the orientation plot for parking the robot from initial configuration of $\mathcal{C}_0=[0.00,1.00,\theta_0]$, where $\theta_0$ is made to vary as; $\theta_0 = 30^\circ$, $\theta_0 = 0^\circ$, and $\theta_0 = -60^\circ$. 
The target configurations were kept the same and given by $\mathcal{C}_f=[2.00,0.00,0.00]$.
As can be seen, in all cases, the robot first converges toward the sliding surface from initial configurations, sliding along this surface until the robot gets closer to the target, and then stabilizes about the same final orientation of $\theta_f = 0^\circ$.

\begin{figure}[t]
\centering
\includegraphics[width=\columnwidth ]{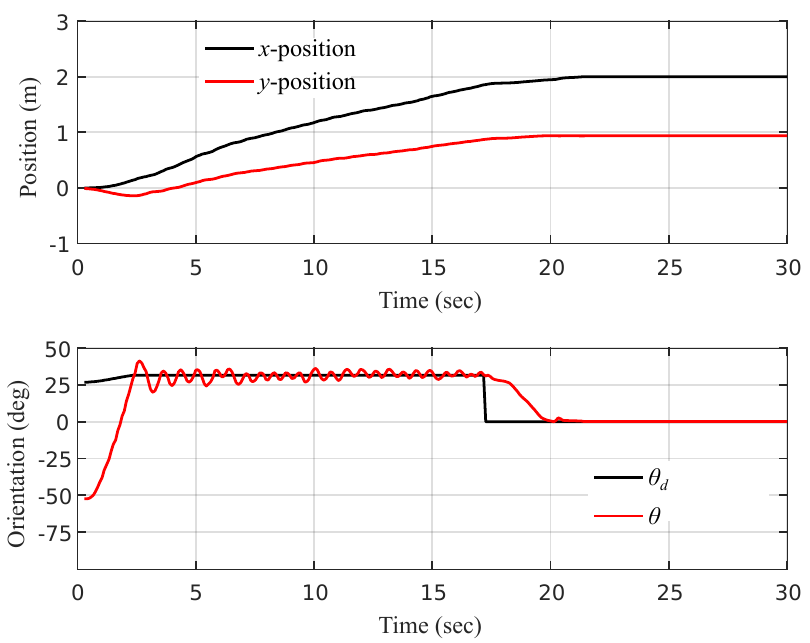} 
\caption{Result for parking the MAMR at $\mathcal{C}_f=[2.00,1.00,0.00]$ from $\mathcal{C}_0=[0.00,0.00,-50.00]$; positions (top) and orientation (bottom). }
\label{arbitrary_parking_xy_0}
\end{figure}

\begin{figure}[t]
\centering
\includegraphics[width=0.8\columnwidth ]{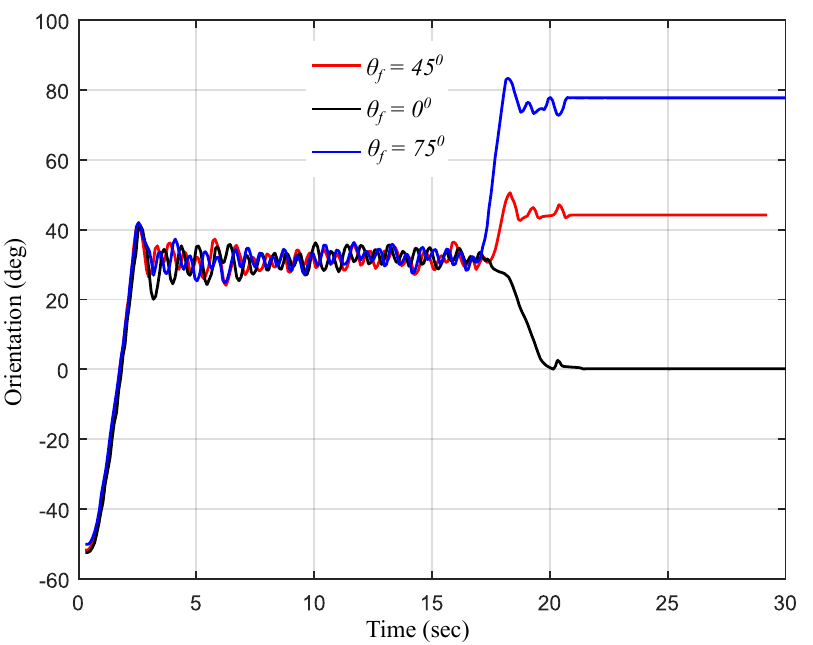} 
\caption{The plot for orientation for parking the robot at $\mathcal{C}_f=[2.00,1.00,\theta_f]$ with $\theta_f$ given by $\theta_f = 0^\circ$, $\theta_f = 45^\circ$, and $\theta_f = 75^\circ$ from initial configuration of $\mathcal{C}_0=[0.00,0.00,-50.00]$.
}
\label{arbitrary_parking_xy_diff_thetad}
\end{figure}

To show that the robot can be arbitrarily parked at any point with any configuration in the capture volume, the target point was moved from the $x$-axis to an arbitrary point in the $x-y$ plane.
The result for parking the robot at final configuration of $\mathcal{C}_f=[2.00,1.00,0.00]$ from an initial configuration of $\mathcal{C}_0=[0.00,0.00,-50.00]$ is shown in Fig.~\ref{arbitrary_parking_xy_0}. 
As can be seen, the robot's $x$ and $y$ positions converge simultaneously to their respective target values.  
Similar to the previous cases, the orientation first converges to the sliding surface, slides along this surface, and then stabilizes to the final constant sliding surface with desired orientation of $\theta_f = 0^\circ$.
Fig.~\ref{arbitrary_parking_xy_diff_thetad} shows the orientation plot for parking the robot to a final configuration of $\mathcal{C}_f=[2.00,1.00,\theta_f]$, where $\theta_f$ is made to vary as; $\theta_f= 0^\circ$, $\theta_f= 45^\circ$, and $\theta_f = 75^\circ$.
The initial configurations were kept constant for all cases and given by $\mathcal{C}_0=[0.00,0.00,-50.00]$.
It can be seen that the orientations converge to the corresponding final values for all test cases.

\begin{figure}[h!]
\centering
\includegraphics[width=0.8\columnwidth ]{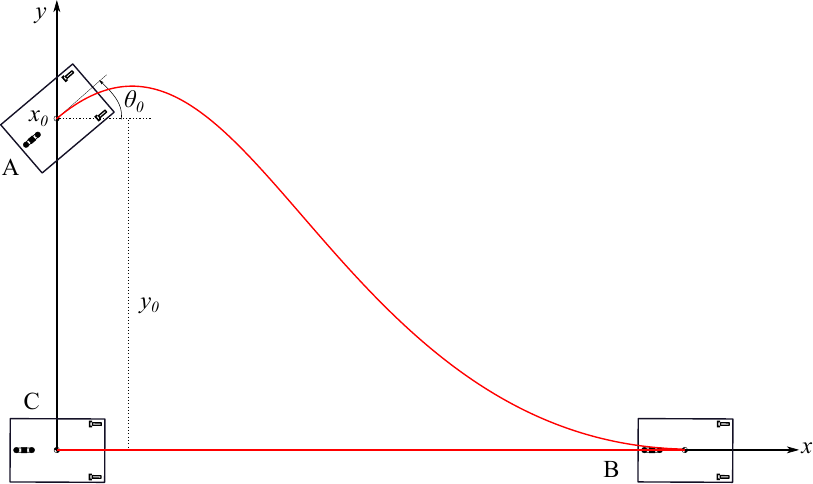} 
\caption{Three point parking problem formulation for the MAMR.}
\label{three_point_parking_problem}
\end{figure}

\begin{figure}[h]
\centering
\includegraphics[width=0.8\columnwidth ]{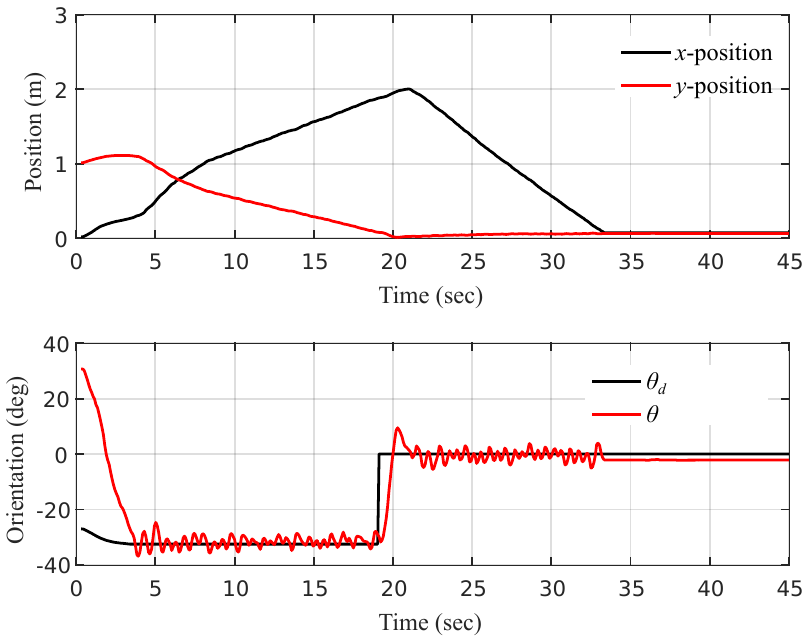} 
\caption{Position and orientation plot for parking the MAMR at $\mathcal{C}_f=[0.00,0.00,0.00]$ from $\mathcal{C}_0=[0.00,1.00,30.00]$ via intermediate configuration defined by $\mathcal{C}_I=[2.00,0.00,-50.00]$.}
\label{three_point_parking}
\end{figure}

Fig.~\ref{three_point_parking_problem} shows a three point parking problem.
The goal is to park the robot at Point $\text{C}$ from Point $\text{ A}$ via Point $\text{B }$. 
In this case, the robot drives forward sliding along the sliding surfaces $S_1(\theta)$ and $S_2(\theta)$ to get Point $\text{B}$.
Then, it drives backward, sliding along $S_2(\theta)$ to get to the final target point ($\text{C}$). 
The experimental results for the positions and orientation to park the robot at Point $\text{C}$ are shown in Fig.~\ref{three_point_parking}. 
The first part (i.e., $\text{A} \rightarrow \text{B}$) is similar to the result shown in Fig. \ref{parking_control_configuration} in which the $x$ position is increasing to $2m$ and the $y$ position is decreasing to $0m$. 
After the intermediate target Point ($\text{B}$) is attained, the robot slides backward by minimizing the error in $x$ position towards $x_f = 0m$ while the $y$ position and the orientation remain as $y_f = 0m$ and $\theta_f = 0^\circ$, respectively.
This shows that the robot can be parked to any desired configuration from a give initial configuration by selecting multiple intermediate target points in between and designing appropriate sliding surfaces between two successive points. 

\section{Conclusions and Recommendations} \label{conclusion_recommendation}

In this work, the effectiveness of the SMC to solve the parking control problem for a MAMR was verified with experiments.
It was shown that the robot can be parking to a desired configuration from a given initial configuration within the capture volume.
The control algorithm first calculates the sliding surfaces (i.e.,~(\ref{slidingDurface_1}) and~(\ref{slidingDurface_2})) which will help drive the robot towards the target point based on the initial configuration.
The dynamics of the robot were exploited to simplify the design of the controllers for both reaching and sliding modes of the SMC.
Under the assumption of sufficiently small velocity at the active brake and small driving force, the dynamics of the robot was approximated as fixed axis rotation.
This simplified the controller design for the reaching phase to a simple stabilizing controller about the desired angle. 
On the sliding surface, because of the fast switching between the two brakes, it was shown that the robot stays on the surface for any driving force for all time.
However, in this application, the robot was stopped when it gets sufficiently close to the target point.
In this case, the robot dynamics were approximated as if it were a one-dimensional mass system, which simplified the controller design.

It was also shown that, for the same initial position, the robot converges to the same sliding surface for different initial orientations. 
This shows that the fixed axis rotation approximation for the reaching phase is successful for the SMC. 
During the reaching phase the desired angle is time varying, resulting in a time varying sliding surface.
However, on the sliding surface, the desired angle remained constant and hence the sliding surface.
In addition, it was shown that the robot can be parked to any final configuration via intermediate target points by appropriately calculating the sliding surfaces between two successive points.
This approach could be used in dealing with obstacle avoidance if the obstacle is fixed and identified beforehand to select an appropriate intermediate target points, or in general for any path following problem.

This work focused on parking the robot from a given initial configuration to some target configuration without attention to rate of convergence and optimality.
In future work, the authors would like to study the optimal solution for the parking problem and use this solution as a bench mark to compare any other sub-optimal solutions. 
The authors also would like to work on the parking problem in the presence of fixed and moving obstacles.
For this case, the multiple point parking scheme using SMC presented in Section \ref{experimental_result} could be exploited. 
Furthermore, one of the main advantages of using the mixed conventional/braking actuation concept over conventional actuators in mobile robotic systems was regaining controllability of the system under primary conventional actuator failure.
Therefore, in future work, we would like to prove this concept and present a supporting examples.



\bibliographystyle{IEEEtran}
\bibliography{MAMR_SMC_1028}

\begin{thebibliography}{10}
\providecommand{\url}[1]{#1}
\csname url@samestyle\endcsname
\providecommand{\newblock}{\relax}
\providecommand{\bibinfo}[2]{#2}
\providecommand{\BIBentrySTDinterwordspacing}{\spaceskip=0pt\relax}
\providecommand{\BIBentryALTinterwordstretchfactor}{4}
\providecommand{\BIBentryALTinterwordspacing}{\spaceskip=\fontdimen2\font plus
\BIBentryALTinterwordstretchfactor\fontdimen3\font minus
  \fontdimen4\font\relax}
\providecommand{\BIBforeignlanguage}[2]{{%
\expandafter\ifx\csname l@#1\endcsname\relax
\typeout{** WARNING: IEEEtran.bst: No hyphenation pattern has been}%
\typeout{** loaded for the language `#1'. Using the pattern for}%
\typeout{** the default language instead.}%
\else
\language=\csname l@#1\endcsname
\fi
#2}}
\providecommand{\BIBdecl}{\relax}
\BIBdecl

\bibitem{schneier2015literature}
M.~Schneier and R.~Bostelman, ``Literature review of mobile robots for
  manufacturing,'' \emph{National Institut of Standards and Technology,
  NewYork}, 2015.

\bibitem{siegwart2011introduction}
R.~Siegwart, I.~R. Nourbakhsh, and D.~Scaramuzza, \emph{Introduction to
  autonomous mobile robots}.\hskip 1em plus 0.5em minus 0.4em\relax MIT press,
  2011.

\bibitem{machado2006overview}
J.~T. Machado and M.~F. Silva, ``An overview of legged robots,'' in
  \emph{International Symposium on Mathematical Methods in Engineering}, 2006.

\bibitem{deshmukh2006robot}
A.~Deshmukh and C.~Amarnath, ``Robot leg mechanisms,'' in \emph{B. Tech.
  Seminar Report, Roll}.\hskip 1em plus 0.5em minus 0.4em\relax Citeseer, 2006.

\bibitem{iverach2012ice}
C.~Iverach-Brereton, A.~Winton, and J.~Baltes, ``Ice skating humanoid robot,''
  in \emph{Conference Towards Autonomous Robotic Systems}.\hskip 1em plus 0.5em
  minus 0.4em\relax Springer, 2012, pp. 209--219.

\bibitem{dudek2005visually}
G.~Dudek, M.~Jenkin, C.~Prahacs, A.~Hogue, J.~Sattar, P.~Giguere, A.~German,
  H.~Liu, S.~Saunderson, A.~Ripsman \emph{et~al.}, ``A visually guided swimming
  robot,'' in \emph{2005 IEEE/RSJ International Conference on Intelligent
  Robots and Systems}.\hskip 1em plus 0.5em minus 0.4em\relax IEEE, 2005, pp.
  3604--3609.

\bibitem{dudek2007aqua}
G.~Dudek, P.~Giguere, C.~Prahacs, S.~Saunderson, J.~Sattar, L.-A.
  Torres-Mendez, M.~Jenkin, A.~German, A.~Hogue, A.~Ripsman \emph{et~al.},
  ``Aqua: An amphibious autonomous robot,'' \emph{AQUA}, vol.~10, p.~43, 2007.

\bibitem{bouabdallah2007design}
S.~Bouabdallah and R.~Siegwart, ``Design and control of a miniature
  quadrotor,'' in \emph{Advances in unmanned aerial vehicles}.\hskip 1em plus
  0.5em minus 0.4em\relax Springer, 2007, pp. 171--210.

\bibitem{cutler2012design}
M.~J. Cutler, ``Design and control of an autonomous variable-pitch quadrotor
  helicopter,'' Ph.D. dissertation, Citeseer, 2012.

\bibitem{stepan2009acroboter}
G.~Stepan, A.~Toth, L.~Kovacs, G.~Bolmsjo, G.~Nikoleris, D.~Surdilovic,
  A.~Conrad, A.~Gasteratos, N.~Kyriakoulis, D.~Chrysostomou \emph{et~al.},
  ``Acroboter: a ceiling based crawling, hoisting and swinging service robot
  platform,'' in \emph{Beyond gray droids: domestic robot design for the 21st
  century workshop at HCI}, vol. 2009, 2009, p.~2.

\bibitem{wang2008biological}
M.~Wang, X.-z. Zang, J.-z. Fan, and J.~Zhao, ``Biological jumping mechanism
  analysis and modeling for frog robot,'' \emph{Journal of Bionic Engineering},
  vol.~5, no.~3, pp. 181--188, 2008.

\bibitem{menon2004gecko}
C.~Menon, M.~Murphy, and M.~Sitti, ``Gecko inspired surface climbing robots,''
  in \emph{Robotics and Biomimetics, 2004. ROBIO 2004. IEEE International
  Conference on}.\hskip 1em plus 0.5em minus 0.4em\relax IEEE, 2004, pp.
  431--436.

\bibitem{morin2008motion}
P.~Morin and C.~Samson, ``Motion control of wheeled mobile robots,'' in
  \emph{Springer Handbook of Robotics}.\hskip 1em plus 0.5em minus 0.4em\relax
  Springer, 2008, pp. 799--826.

\bibitem{de2001control}
A.~De~Luca, G.~Oriolo, and M.~Vendittelli, ``Control of wheeled mobile robots:
  An experimental overview,'' in \emph{Ramsete}.\hskip 1em plus 0.5em minus
  0.4em\relax Springer, 2001, pp. 181--226.

\bibitem{Walelign2018_SMC_FLC}
W.~Nikshi, R.~Hoover, M.~Bedillion, and J.~Simmons, ``Fuzzy logic control for
  mixed conventional/braking actuation mobile robots,'' \emph{IEEE Transactions
  on Systems, Man, and Cybernetics, Part A: Systems (Under Review)}, 2018.

\bibitem{iagnemma2000mobile}
K.~Iagnemma, A.~Rzepniewski, S.~Dubowsky, P.~Pirjanian, T.~Huntsberger, and
  P.~Schenker, ``Mobile robot kinematic reconfigurability for rough-terrain,''
  in \emph{Proc. SPIE}, vol. 4196, 2000, pp. 413--420.

\bibitem{iagnemma2003control}
K.~Iagnemma, A.~Rzepniewski, S.~Dubowsky, and P.~Schenker, ``Control of robotic
  vehicles with actively articulated suspensions in rough terrain,''
  \emph{Autonomous Robots}, vol.~14, no.~1, pp. 5--16, 2003.

\bibitem{Walelign2017SMC}
W.~Nikshi, M.~Bedillion, and R.~Hoover, ``Nonlinear control synthesis for
  parking control of mixed conventional/braking actuation mobile robots,'' in
  \emph{2017 American Control Conference}.\hskip 1em plus 0.5em minus
  0.4em\relax IEEE, 2017.

\bibitem{simmons2016mechatronic}
J.~W. Simmons, W.~M. Nikshi, M.~D. Bedillion, and R.~C. Hoover, ``Mechatronic
  design of a mixed conventional/braking actuation mobile robot,'' in
  \emph{ASME 2016 International Mechanical Engineering Congress and
  Exposition}.\hskip 1em plus 0.5em minus 0.4em\relax American Society of
  Mechanical Engineers, 2016.

\bibitem{nikshi2016parking}
W.~M. Nikshi, M.~D. Bedillion, and R.~C. Hoover, ``Parking control of mixed
  conventional/braking actuation mobile robots using fuzzy logic control,'' in
  \emph{ASME, International Mechanical Engineering Congress and
  Exposition}.\hskip 1em plus 0.5em minus 0.4em\relax American Society of
  Mechanical Engineers, 2016.

\bibitem{solea2009sliding}
R.~Solea, A.~Filipescu, and U.~Nunes, ``Sliding-mode control for
  trajectory-tracking of a wheeled mobile robot in presence of uncertainties,''
  in \emph{Asian Control Conference, 2009. ASCC 2009. 7th}.\hskip 1em plus
  0.5em minus 0.4em\relax IEEE, 2009, pp. 1701--1706.

\bibitem{tzafestas2013introduction}
S.~G. Tzafestas, \emph{Introduction to mobile robot control}.\hskip 1em plus
  0.5em minus 0.4em\relax Elsevier, 2013.

\bibitem{chang2002design}
S.-J. Chang and T.-H.~S. Li, ``Design and implementation of fuzzy
  parallel-parking control for a car-type mobile robot,'' \emph{Journal of
  Intelligent \& Robotic Systems}, vol.~34, no.~2, pp. 175--194, 2002.

\bibitem{sira1987variable}
H.~SIRA-RAMIREZ, ``Variable structure control of non-linear systems,''
  \emph{International journal of systems science}, vol.~18, no.~9, pp.
  1673--1689, 1987.

\end{thebibliography}

\end{document}